\documentclass[aps,prl,reprint,superscriptaddress]{revtex4-2}
\usepackage[colorlinks,linkcolor=blue,anchorcolor=blue,citecolor=blue]{hyperref}
\usepackage{graphicx}
\usepackage{amsmath}
\usepackage{braket}

\begin{document}

\title{Coherence Spreading Law in Quenched Homogeneous Bose Gases}
\author{Yi Cao}
\email{caoyi24@gscaep.ac.cn}
\affiliation{Graduate School, China Academy of Engineering Physics, Beijing 100193, China}
\author{Li-Chen Zhao}
\affiliation{School of Physics, Northwest University, Xi'an, 710127, China}
\affiliation{NSFC-SPTP Peng Huanwu Center for Fundamental Theory, 710127, Xi'an, China}
\affiliation{Shaanxi Key Laboratory for Theoretical Physics Frontiers, 710127, Xi'an, China}
\affiliation{Fundamental Discipline Research Center for Quantum Science and technology of Shaanxi Province, 710127, Xi'an, China}
\author{Jie Liu}
\email{jliu@gscaep.ac.cn}
\affiliation{Graduate School, China Academy of Engineering Physics, Beijing 100193, China}
\date{\today}

\begin{abstract}
We investigate nonequilibrium coherence spreading in quenched homogeneous bosonic gases both analytically and numerically. Our ensemble-averaged Gross-Pitaevskii trajectory approach and kinetic analysis confirm the linear-in-time growth of the squared coherence length, and explicitly give the spreading-law dependence on the ratio of initial energy to healing energy. When the ratio is relatively small, the squared coherence length grows at a universal rate determined by the ratio of Planck's constant to the particle mass, insensitive to interaction strength and initial momentum distributions. However, our further calculations uncover fundamental limitations to such universality and establish its phase boundaries governed by low system energies, weak interparticle interactions and many-body integrability. Our theory fully reproduces the key experimental signatures and resolves the puzzle of the dynamical origin of the recently observed universal coherence-spreading speed limit [Nature 647, 608-612 (2025)]. Going beyond existing experimental observations, we provide some testable predictions for ultracold-atom experiments and clarify the conditions for the emergence of universal coherent dynamics in nonequilibrium bosonic many-body systems.
\end{abstract}

\maketitle

How isolated many-body quantum systems depart from their initial states and relax toward equilibrium is a core question in nonequilibrium statistical physics~\cite{Gemmer2009,DeutschPRA1991, SrednickiPRE1994, ReimannPRL2008, RigolNature2008, PolkovnikovRMP2011, EisertNP2015, DAlessioAP2016}. The formation of Bose-Einstein condensates (BECs) out of far-from-equilibrium ultracold bosonic gases provides a paradigmatic arena for exploring the crossover between equilibrium statistical mechanics and genuine quantum many-body relaxation~\cite{GardinerPRL1997, DalfovoRMP1999, BijlsmaPRA2000, DavisPRA2000, Pitaevskii2016}.
Kinetic descriptions of weakly interacting bosonic gases have highlighted the intertwined processes of occupation redistribution, quasicondensate formation, and the eventual emergence of long-range phase coherence~\cite{kaganJETP1992, Kagan1995, ProukakisJPB2008, Griffin2009}. The growth of a correlation length and the associated dynamical scaling are central concepts in the theory of nonequilibrium critical dynamics~\cite{HohenbergRMP1977, NavonSci2015}. In addition, far-from-equilibrium bosonic gases also exhibit prethermal regimes and nonthermal scaling phenomena~\cite{BergesPRL2004, NowakPRA2012, PruferNat2018}. The underlying dynamics involves not only energy thermalization but also quantum phase-driven effects, including coherence spreading~\cite{MartirosyanNat2025} and phase ordering~\cite{DamlePRA1996}.

With their exceptional tunability, ultracold atomic systems make these nonequilibrium effects experimentally accessible~\cite{BlochRMP2008, KinoshitaNat2006, LiuNJP2008, LiuPRA2010, TrotzkyNP2012, EigenNat2018, Liu2018}. In the recent experiments of Ref.~\cite{MartirosyanNat2025}, isolated homogeneous atomic gases are prepared in isotropic, incoherent, far-from-equilibrium low-energy states. The interatomic interactions are then quenched from zero to a finite value via a Feshbach resonance. The ensuing evolution corresponds to nonequilibrium relaxation of an isolated interacting many-body system: scattering among the initially populated momentum modes redistributes their occupations and builds up coherence as the gas evolves toward equilibrium. The coherence-spreading speed is measured and found to be a universal value, which is largely insensitive to the interaction strength. These experimental observations of bosonic coherence spreading are remarkable and serve as valuable benchmarks for theories of far-from-equilibrium universality. Nevertheless, the underlying microscopic mechanisms remain incompletely understood.

In this Letter, we deal with the quantum many-body relaxation process using an ensemble-averaged Gross-Pitaevskii (GP) trajectory approach and weak-wave kinetic theory. We have derived an analytical solution for coherence spreading, unraveled the emergence of the universal coherence dynamics and identified inherent bounds to the observed universality. Our theory uncovers that the spreading rate intrinsically depends on the ratio of initial energy to healing energy and is strictly governed by the system's integrability, predicting a quantum-thermal dynamical crossover below the BEC critical temperature. 

The experiment reported in Ref.~\cite{MartirosyanNat2025} explores a dilute $^{39}\mathrm{K}$ Bose gas in a regime of low-energy highly occupied modes, where phase-space techniques based on the Wigner representation offer an approximate description of relaxation toward equilibrium. In this situation, quantum field evolution can be cast into equations of motion resembling the GP equation, yet augmented by stochastic corrections that capture quantum effects within a controlled approximation~\cite{BlakieAP2008}. In this spirit, we develop an ensemble-averaged GP trajectory approach. For each trajectory $\alpha$, we sample the initial mode occupations $n_{\mathbf{k}}(0)$ and phases $\theta_{\mathbf{k}}(0)$ from $P(\{n_{\mathbf{k}}(0)\},\{\theta_{\mathbf{k}}(0)\})=P_{n}(\{n_{\mathbf{k}}(0)\})\prod_{\mathbf{k}}(2\pi)^{-1}$, where $P_{n}$ arises from Gaussian sampling of the complex mode amplitudes consistent with the prescribed momentum distribution. Each realization subsequently evolves deterministically according to $i\hbar\partial_{t}\psi_{\alpha}=[-\hbar^{2}\nabla_{\mathbf{r}}^{2}/2m+(4\pi\hbar^{2}a_{s}/m)|\psi_{\alpha}|^{2}]\psi_{\alpha}$, where $m$ is the atomic mass and $a_{s}$ is the post-quench $s$-wave scattering length. The trajectories are propagated in a periodic cubic box of volume $V=L^{3}$, with fixed mean density $n=N/V$ and fixed particle number $N\equiv\int_{V}d^{3}\mathbf{r}|\psi_{\alpha}(\mathbf{r},t)|^{2}$ for each realization. As in the experiment, we parameterize the initial state by its pre-quench mean single-particle kinetic energy $E_{\mathrm{kin}}^{\mathrm{pre}}=k_{\mathrm{B}}T$, where $k_{\mathrm{B}}$ is the Boltzmann constant. To vary the initial energy, we rescale the experimental $T=20\mathrm{nK}$ momentum distribution to match the target kinetic energy and renormalize it to preserve particle number. For an observable $A$, we denote its value along trajectory $\alpha$ by $A_{\alpha}(t)$ and define the ensemble average over $M=32$ trajectories per protocol as $\langle A(t)\rangle\equiv M^{-1}\sum_{\alpha=1}^{M}A_{\alpha}(t)$. For a given trajectory, the normalized first-order correlation function~\cite{NaraschewskiPRA1999} $g_{\alpha}^{(1)}(\mathbf{R},t)\equiv N^{-1}\int_{V}d^{3}\mathbf{r}\psi_{\alpha}^{\ast}(\mathbf{r}+\mathbf{R},t)\psi_{\alpha}(\mathbf{r},t)$ obeys $\int_{V}d^{3}\mathbf{R}g_{\alpha}^{(1)}(\mathbf{R},t)=Vn_{0,\alpha}(t)$, where $n_{0,\alpha}(t)\equiv(NV)^{-1}\left|\int_{V}d^{3}\mathbf{r}\psi_{\alpha}(\mathbf{r},t)\right|^{2}$ is the zero-momentum fraction. The correlation volume for a single trajectory reads $\ell_{\alpha}^{3}(t)=\int_{V}d^{3}\mathbf{R}g_{\alpha}^{(1)}(\mathbf{R},t)=Vn_{0,\alpha}(t)$. We define a normalized coherence length $\ell(t)$ by scaling the ensemble-averaged correlation volume relative to its equilibrium value. This definition yields $\ell^{2}(t)\equiv L^{2}\left[\langle n_{0}(t)\rangle/\langle n_{0}^{\mathrm{eq}}\rangle\right]^{2/3}$, where $n_{0}^{\mathrm{eq}}$ is the saturated condensation fraction of the ultracold Bose atoms.
\begin{figure}[t]
\centering
\includegraphics[width=1\linewidth]{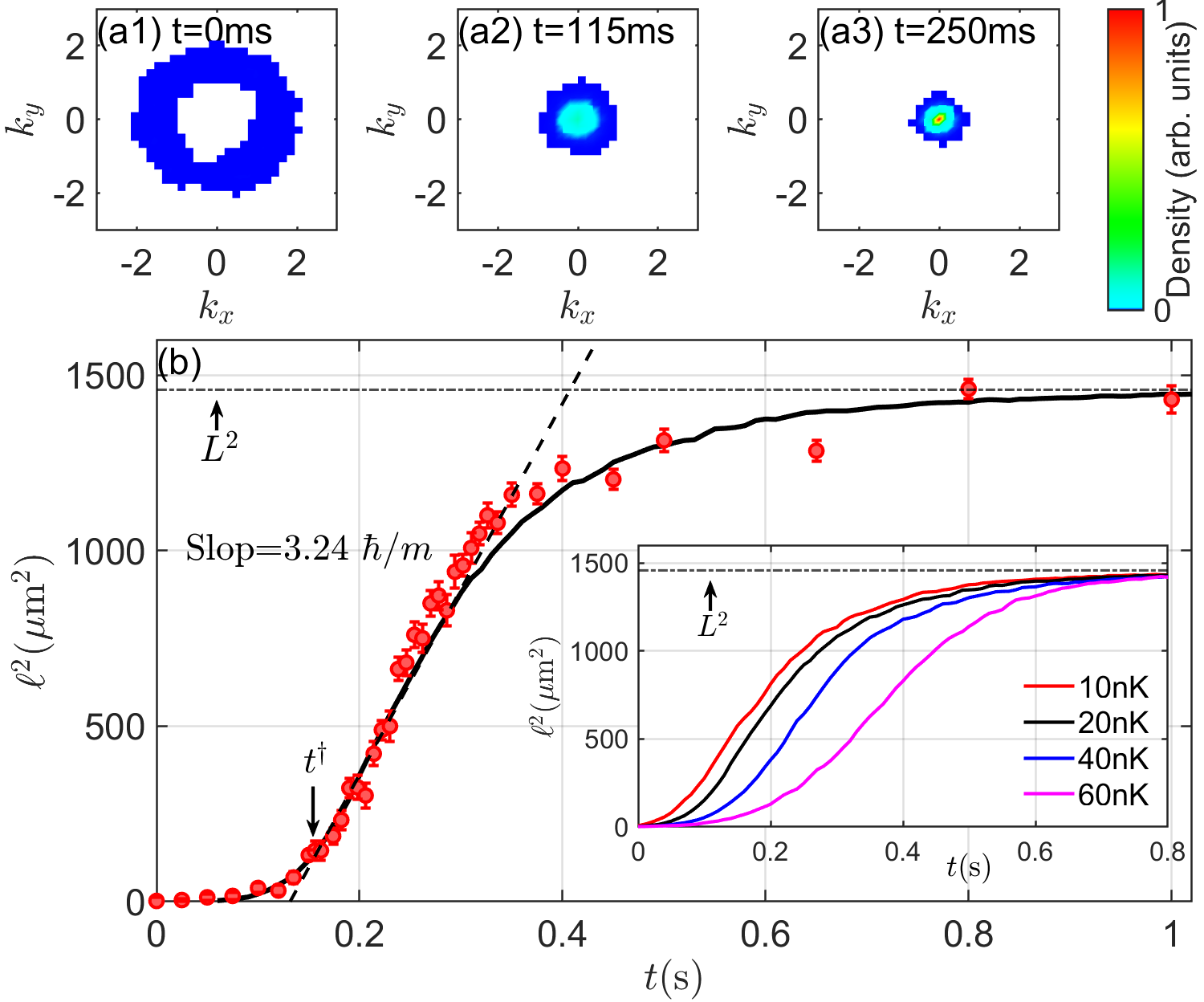}
\caption{Time evolution of the $k_{z}=0$ momentum-space slice (a) and coherence spreading (b) for the $P_{1}$ initial-state protocol from Ref.~\cite{MartirosyanNat2025} at $T=20{\mathrm{nK}}$. Inset in (b): coherence spreading for different initial energies. The simulations are carried out for $^{39}\mathrm{K}$ atoms with $n=5.4\mu\mathrm{m}^{-3}$, $a_{s}=100a_{0}$, and $V=5.54\times10^{4}\mu\mathrm{m}^{3}$, where $a_{0}$ is the Bohr radius.}
\label{Figure1}
\end{figure}

\begin{figure}[t]
\centering
\includegraphics[width=1\linewidth]{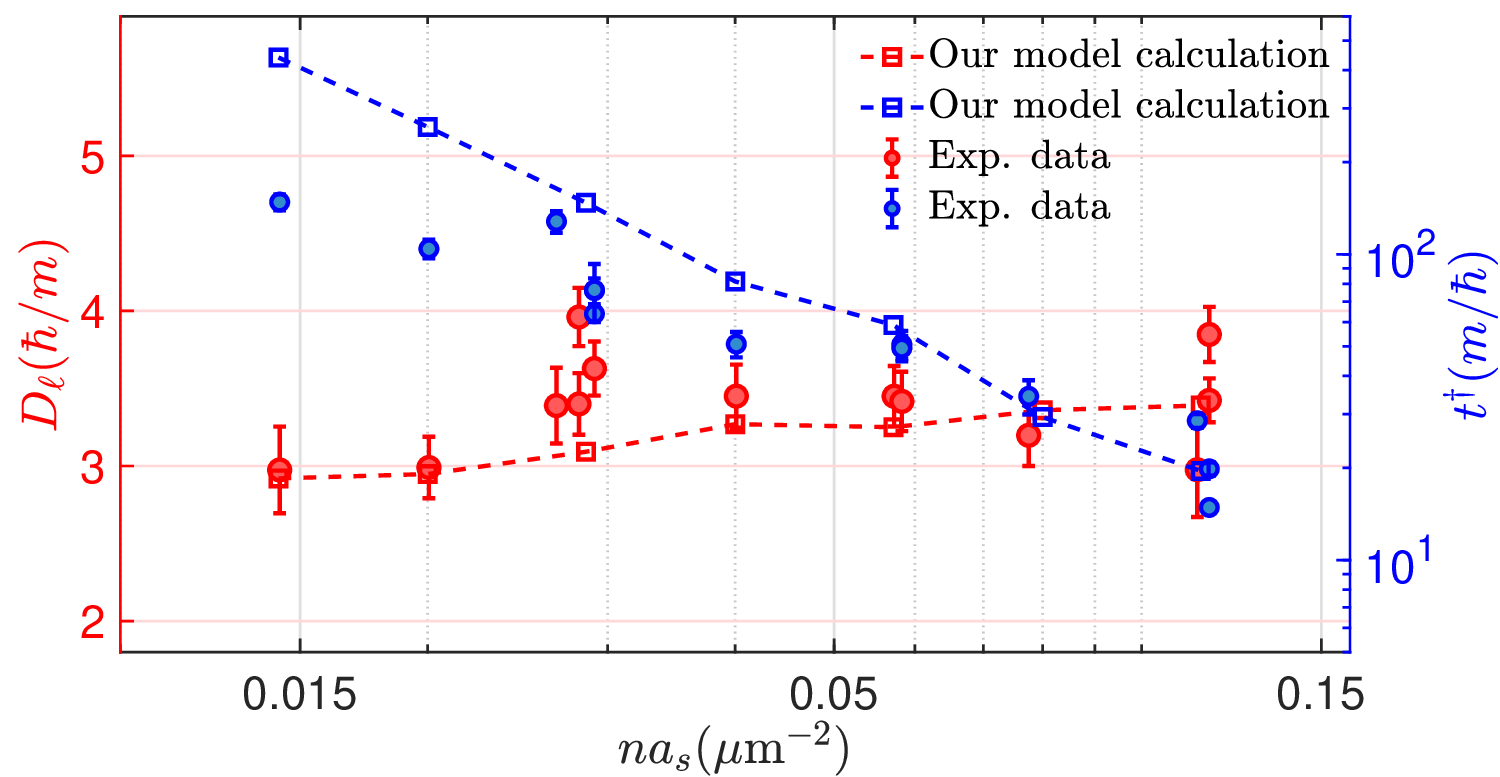}
\caption{Coherence-spreading rate $D_{\ell}$ (red) and energy pre-equilibration time $t^{\dagger}$ (blue) versus interaction strength at $T=20\mathrm{nK}$, with all other parameters as in Fig.~\ref{Figure1}. The blue squares use the $P_{1}$ protocol, while the red squares show the mean across the three initial-state protocols of Ref.~\cite{MartirosyanNat2025}.}
\label{Figure2}
\end{figure}

In Fig.~\ref{Figure1}, we present numerical simulations of the time evolution for both momentum-space relaxation and coherence spreading, which are in excellent agreement with the experiments reported in Ref.~\cite{MartirosyanNat2025}. We perform extensive numerical simulations for different initial energies, initial-momentum distributions, and interaction strengths. These simulations show that relaxation toward equilibrium comprises three distinct stages: energy pre-equilibration, coherence spreading, and equilibrium with a saturated coherence length. In the first stage, through energy relaxation, the ultracold atoms attain a Rayleigh-Jeans (RJ) distribution~\cite{Nazarenko2011}, and the characteristic time scale $t^{\dagger}$ strongly depends on the initial energy, initial-momentum distribution, and interaction strength~\cite{Lifshitz1981}. For $T=20\mathrm{nK}$ in Fig.~\ref{Figure2}, our model predicts $t^{\dagger}\propto(na_{s})^{-1.1(1)}$, whereas the experiment reports $t^{\dagger}\propto(na_{s})^{-0.9(1)}$. Both results point to the scaling $t^{\dagger}\propto(na_{s})^{-1}$. In the coherence-spreading stage, the system exhibits diffcusion-like behavior, that is, the square of the coherence length scales linearly with time. We define the coherence-spreading rate as $D_{\ell}=d\ell^{2}/dt$. Of particular interest, this growth rate exhibits universal behavior: it is independent of both the initial-momentum distribution and the interaction strength, as shown in Fig.~\ref{Figure2}. Our model calculations give the speed of coherence spreading to be $3.2\pm 0.2 \hbar/m$ approximately. Yet, while the coherence-spreading behavior for this universal growth rate has been anticipated, our further numerical findings reveal that its universality is limited: as the initial energy increases, we find that the growth rate decays markedly (see the inset in Fig.~\ref{Figure1}).

To gain insight into the dynamical origin and physical mechanisms behind these nontrivial coherence spreading behaviors, we introduce the dimensionless variables $\mathbf{x}=\mathbf{r}/\xi$, $\tau=t/t_{\xi}$, and $\phi=\psi/\sqrt{n}$, where $\xi=(8\pi na_{s})^{-1/2}$ is the healing length and $t_{\xi}=m\xi^{2}/\hbar$ is the corresponding time scale. The cubic box has dimensionless side length $\mathcal{L}=L/\xi$ and volume $\mathcal{V}=\mathcal{L}^{3}$, with the field normalized as $\mathcal{V}^{-1}\int_{\mathcal{V}}d^{3}\mathbf{x}|\phi|^{2}=1$. The GP equation then takes the simple form $i\partial_{\tau}\phi=\left(-\nabla_{\mathbf{x}}^{2}/2+|\phi|^{2}/2\right)\phi$.

Choose a reference time $\tau_{i}$ after energy pre-equilibration but before the coherence length approaches the box size, for example $\tau_{i}=t^{\dagger}/t_{\xi}$, and set $\Delta\tau=\tau-\tau_{i}$. In this regime, the ensemble-averaged momentum distribution $\langle n_{\mathbf{k}}(\tau_{i}+\Delta\tau)\rangle$ consists of a narrow coherent peak with characteristic width $k_{\mathrm{coh}}$ and a broad background with width $k_{\mathrm{bg}}$. When $k_{\mathrm{coh}}\ll k_{\ast}\ll k_{\mathrm{bg}}$, the intermediate momentum $k_{\ast}$ separates these two sectors. We define the total weight and normalized real-space profile of the low-momentum sector as $Z(\Delta\tau)=\sum_{|\mathbf{k}|<k_{\ast}}\langle n_{\mathbf{k}}(\tau_{i}+\Delta\tau)\rangle$ and $h(\mathbf{x},\Delta\tau)=Z^{-1}(\Delta\tau)\sum_{|\mathbf{k}|<k_{\ast}}\langle n_{\mathbf{k}}(\tau_{i}+\Delta\tau)\rangle e^{i\mathbf{k}\cdot\mathbf{x}}$, respectively, so that $h(\mathbf{0},\Delta\tau)=1$. The corresponding ensemble-averaged correlation function $g^{(1)}\equiv\langle g_{\alpha}^{(1)}\rangle$ decomposes into the complementary background and low-momentum contributions as
\begin{equation}\label{Eq1}
    g^{(1)}(\mathbf{x},\tau_{i}+\Delta\tau)=g_{\mathrm{bg}}^{(1)}(\mathbf{x},\Delta\tau)+Z(\Delta\tau)h(\mathbf{x},\Delta\tau).
\end{equation}
The broad background term does not directly contribute the coherence spreading because $g_{\mathrm{bg}}^{(1)}$ contains no zero-momentum component and Fourier orthogonality gives $\int_{\mathcal{V}}d^{3}\mathbf{x}g_{\mathrm{bg}}^{(1)}(\mathbf{x},\Delta\tau)=0$. The second term of $Zh$ directly contributes to the correlation volume that defines $\ell$, with the background influencing $Z$ and $h$ via particle exchange and phase fluctuations.

After energy pre-equilibration, we describe particle exchange between the low-momentum sector $k<k_{\ast}$ and the background using weak-wave kinetics~\cite{DyachenkoPhyD1992, ZhuPRE2022}. This exchange occurs through four-wave scattering that conserves total momentum and energy. We assume that the modes dominating these processes remain close to a quasistationary RJ distribution. Particle exchange between the two sectors is then nearly balanced, resulting in a negligible net flux across $k_{\ast}$, i.e., $Z(\Delta\tau)\simeq Z(0)$ (see End Matter). However, particles can still redistribute within the low-momentum sector, narrowing the coherent peak and increasing the zero-mode occupation.

We next determine the evolution of $h$ using the normalized kernel $K(\mathbf{x},\Delta\tau)=h(\mathbf{x},\Delta\tau)/\int d^{3}\mathbf{x}h(\mathbf{x},\Delta\tau)$, with Fourier transform $\widetilde{K}(\mathbf{k},\Delta\tau)=\widetilde{h}(\mathbf{k},\Delta\tau)/\widetilde{h}(\mathbf{0},\Delta\tau)$. We describe its broadening by the coarse-grained relaxation equation $\partial_{\Delta\tau}\widetilde{K}=-\Gamma_{k}(\Delta\tau)\widetilde{K}$, where $\Gamma_{k}$ is the effective relaxation rate and normalization fixes $\Gamma_{0}=0$~\cite{HohenbergRMP1977}. The isotropy of the gas makes $\Gamma_{k}$ independent of direction, and analyticity at $k=0$ permits an expansion in even powers, giving $\Gamma_{k}=D_{\phi}(\Delta\tau)k^{2}+O(k^{4})$, where $D_{\phi}$ is the dimensionless diffusion coefficient. The relaxation equation then gives $\widetilde{K}(\mathbf{k},\Delta\tau)\simeq e^{-\sigma(\Delta\tau)k^{2}}$ for the reference-time profile $\widetilde{K}(\mathbf{k},0)\simeq e^{-\sigma_{i}k^{2}}$, where $\sigma(\Delta\tau)=\sigma_{i}+\int_{0}^{\Delta\tau}dwD_{\phi}(w)$. In the continuum approximation, we take the inverse Fourier transform and use $h(\mathbf{0},\Delta\tau)=1$ to obtain $h(\mathbf{x},\Delta\tau)\simeq e^{-x^{2}/(4\sigma)}$, where $x=|\mathbf{x}|$. Thus $\sqrt{\sigma}$ characterizes the spatial width of $h$, and $\dot{\sigma}=D_{\phi}$, with a dot denoting $\partial_{\Delta\tau}$.

We estimate $D_{\phi}$ from finite-memory transfer involving the nonequilibrium coherent peak. With $\bar{n}_{j}=\langle n_{j}\rangle$ and $\mathcal{I}_{j}=\partial_{\Delta\tau}\bar{n}_{j}$, using $\widetilde{K}=\bar{n}_{\mathbf{k}}/\bar{n}_{\mathbf{0}}$, we obtain $\Gamma_{k}=\mathcal{I}_{\mathbf{0}}/\bar{n}_{\mathbf{0}}-\mathcal{I}_{\mathbf{k}}/\bar{n}_{\mathbf{k}}$. Two-body scattering transfers particles from modes $\mathbf{k}$ and $\mathbf{p}$ to modes $\mathbf{q}$ and $\mathbf{r}$, conserving momentum $\mathbf{k}+\mathbf{p}=\mathbf{q}+\mathbf{r}$, with field expansion $\phi=\sum_{\mathbf {k}}c_{\mathbf {k}}e^{i\mathbf {kx}}$. Within the weak-wave kinetic approximation~\cite{DyachenkoPhyD1992}, we use the GP equation and retain only products of three two-point averages to obtain $(\partial_{\tau}-i\Delta\omega)Q\simeq i\mathcal{F}$~\cite{AndradeFluids2019}. Here $Q=\langle c_{\mathbf{k}}^{\ast}c_{\mathbf{p}}^{\ast}c_{\mathbf{q}}c_{\mathbf{r}}\rangle_{\mathrm{c}}$ is the connected four-point correlation (fourth cumulant), $\Delta\omega=\omega_{k}+\omega_{p}-\omega_{q}-\omega_{r}$, and $\mathcal{F}_{\mathbf{kpqr}}=\bar{n}_{\mathbf{k}}\bar{n}_{\mathbf{p}}\bar{n}_{\mathbf{q}}\bar{n}_{\mathbf{r}}(1/\bar{n}_{\mathbf{k}}+1/\bar{n}_{\mathbf{p}}-1/\bar{n}_{\mathbf{q}}-1/\bar{n}_{\mathbf{r}})$ is the occupation source~\cite{ZhuPRE2022}. Since the disconnected pairings in the four-point average are real and do not enter the imaginary part, differentiating $n_{\mathbf{k}}=|c_{\mathbf{k}}|^{2}$ gives $\mathcal{I}_{\mathbf{k}}=2\operatorname{Re}\langle c_{\mathbf{k}}^{\ast}\partial_{\tau}c_{\mathbf{k}}\rangle=\sum_{\mathbf{pqr}}\delta_{\mathbf{k}+\mathbf{p},\mathbf{q}+\mathbf{r}}\operatorname{Im}Q_{\mathbf{kpqr}}$. We focus on the intermediate spreading regime, where the initial term $e^{i\Delta\omega\Delta\tau}Q(\tau_{i})$ in the solution of $Q$ has dephased in the channel-averaged particle transfer. Occupation factors are evaluated at the current time because channel dephasing is faster than particle redistribution~\cite{ZhuPRE2022, NewellARFM2011, BhongalePRA2002}. The source-generated part of $Q$ is $i\mathcal{F}\int_{0}^{\Delta\tau}dse^{i\Delta\omega s}$, whose imaginary part gives $\mathcal{I}_{\mathbf{k}}\simeq\sum_{\mathbf{pqr}}\delta_{\mathbf{k}+\mathbf{p},\mathbf{q}+\mathbf{r}}\mathcal{F}_{\mathbf{kpqr}}\int_{0}^{\Delta\tau}ds\cos(\Delta\omega s)$~\cite{DyachenkoPhyD1992}. On resonance, $\mathcal{F}$ vanishes for a full RJ distribution, while departures of the coherent peak from RJ drive the transfer. We define $\mathcal{R}_{k}(s)$ as the integrand of $\mathcal{I}_{\mathbf{0}}/\bar{n}_{\mathbf{0}}$ minus that of $\mathcal{I}_{\mathbf{k}}/\bar{n}_{\mathbf{k}}$, giving $\Gamma_{k}\simeq\int_{0}^{\Delta\tau}ds\mathcal{R}_{k}(s)$. Isotropy and analyticity at $\mathbf{k}=0$, together with $\mathcal{R}_{0}(s)=0$, give $\mathcal{R}_{k}(s)=Ak^{2}f(s)+O(k^{4})$, where $f(0)=1$. Comparing the $k^{2}$ terms gives $D_{\phi}(\Delta\tau)\simeq A\int_{0}^{\Delta\tau}f(s)ds$. For $k\ll k_{\mathrm{bg}}$, we estimate $f(s)$ from the $\mathbf{k}=0$ phase factor, where momentum conservation gives $\mathbf{p}=\mathbf{q}+\mathbf{r}$ and $\Delta\omega_{\mathbf{0pqr}}=\mathbf{q}\cdot\mathbf{r}$. Approximating the occupation-dependent channel weights by independent Gaussians with a characteristic width $k_{\mathrm{bg}}$ in each momentum direction gives $f(s)\simeq\langle\cos(\mathbf{q}\cdot\mathbf{r}s)\rangle=(1+k_{\mathrm{bg}}^{4}s^{2})^{-3/2}$, with the effective correlation time defined as $\tau_{c}\equiv\int_{0}^{\infty}f(s)ds=k_{\mathrm{bg}}^{-2}$. Estimating the equal-time value $\mathcal{R}_{k}(0)$ by $\omega_{b}^{2}$ and its momentum-variation scale by $k_{\mathrm{bg}}$ gives $\mathcal{R}_{k}(0)\simeq\omega_{b}^{2}(k/k_{\mathrm{bg}})^{2}$ and $A=\lim_{k\to0}\mathcal{R}_{k}(0)/k^{2}\simeq\omega_{b}^{2}/k_{\mathrm{bg}}^{2}=k_{\mathrm{bg}}^{2}/4$, where $\omega_{b}=k_{\mathrm{bg}}^{2}/2$. We describe the correlation decay by $f(s)\simeq e^{-s/\tau_{c}}$, matching the initial value and integrated area~\cite{LeiPNAS2016,OttobreNon2011}, yielding $D_{\phi}(\Delta\tau)\simeq A\tau_{c}(1-e^{-\Delta\tau/\tau_{c}})\simeq\frac{1}{4}(1-e^{-\Delta\tau/\tau_{c}})$.

Integrating $\dot{\sigma}=D_{\phi}$ gives $\sigma(\Delta\tau)=\sigma_{i}+\{\Delta\tau-\tau_{c}[1-e^{-\Delta\tau/\tau_{c}}]\}/4$. Using Eq.~\eqref{Eq1} and the definition of the coherence length, $\ell^{2}(\tau_{i}+\Delta\tau)/\xi^{2}\simeq4\pi\sigma[Z(0)/\langle n_{0}^{\mathrm{eq}}\rangle]^{2/3}$, differentiating with respect to physical time and using $\xi^{2}/t_{\xi}=\hbar/m$, we finally obtain
\begin{equation}\label{Eq2}
    D_{\ell}(\Delta\tau)\simeq\pi\frac{\hbar}{m}\left[\frac{Z(0)}{\langle n_{0}^{\mathrm{eq}}\rangle}\right]^{2/3}\left(1-e^{-\Delta\tau/\tau_{c}}\right).
\end{equation}
When the system's temperature is much lower than the BEC critical temperature $T_{C}$, the momentum distribution will approach a delta function, i.e., $Z(0)\simeq\langle n_{0}^{\mathrm{eq}}\rangle$. In this situation, Eq.~\eqref{Eq2} then reduces to a very simple form of $D_{\ell}\simeq\pi\hbar/m$. Here, we have assumed the memory time $\tau_{c}$ is much smaller than our evolution time~\cite{BhongalePRA2002}. This memory time characterizes the decay of scattering correlations in the noncondensed background through channel dephasing.

\begin{figure}[t]
\centering
\includegraphics[width=1\linewidth]{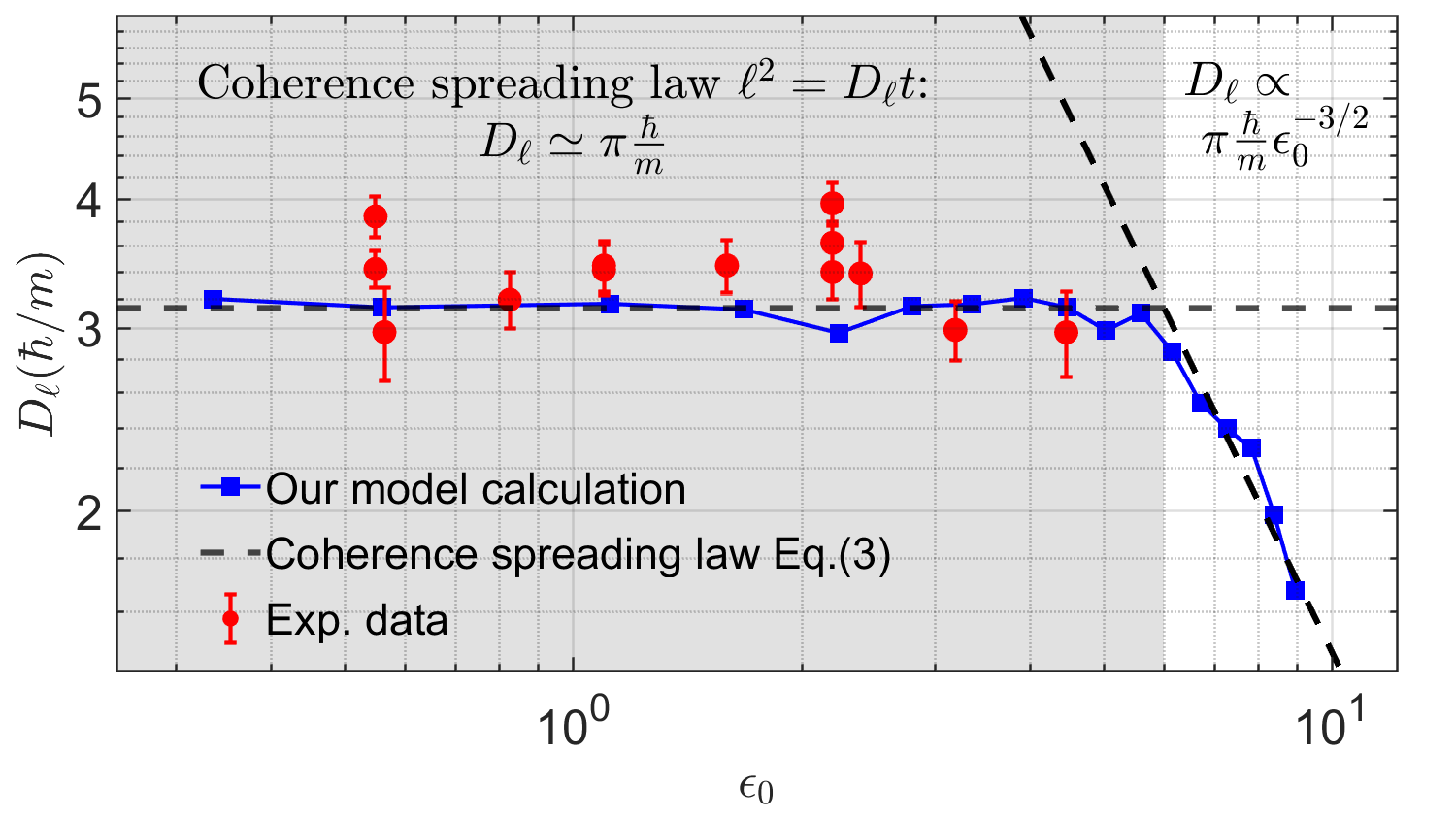}
\caption{Coherence-spreading rate $D_{\ell}$ versus the ratio of initial energy to healing energy. The simulations are carried out by varying the initial energy, with all other parameters identical to those in Fig.~\ref{Figure1}.}
\label{Figure3}
\end{figure}

$D_{\ell}\propto\hbar/m$ is the so-called coherence-spreading speed limit, first reported in the recent experiment of Ref.~\cite{MartirosyanNat2025}. Our weak-wave kinetic theory reproduces this experimental observation and identifies the precise coefficient $\pi$.
In the dimensionless GP equation, the energy unit is the healing energy of $E_{\xi}=\hbar^{2}/(m\xi^{2})$. We find that the coherence spreading speeds are solely determined by the parameter of scaled initial kinetic energy, i.e., $\epsilon_{0}\equiv E_{\mathrm{kin}}^{\mathrm{pre}}/E_{\xi}$. Our numerical results further confirm that $D_{\ell}$ remains approximately constant over an extensive parameter range. However, when $\epsilon_{0}$ exceeds $\epsilon^{\ast}\simeq 6$, $D_{\ell}$ is no longer constant and instead follows a power-law decay with increasing $\epsilon_{0}$. This indicates that coherence spreading speed in this regime is suppressed and depends explicitly on the interaction strength. This behavior is illustrated in Fig.~\ref{Figure3}, where $D_{\ell}$ is given by
\begin{equation}
    D_{\ell}\simeq\begin{cases}
        \pi\frac{\hbar}{m}, & \epsilon_{0}\le\epsilon^{\ast},\\
        \pi\frac{\hbar}{m}\left({\frac{\epsilon_{0}}{\epsilon^{\ast}}}\right)^{-\frac{3}{2}}, & \epsilon_{0}>\epsilon^{\ast}.
    \end{cases}
\end{equation}
For the experimental parameters $n=5.4\mu \mathrm{m}^{-3}$ and $a_{s}=100a_{0}$~\cite{MartirosyanNat2025}, the energy corresponding to $\epsilon^{\ast}$ is $E_{\mathrm{kin}}^{\mathrm{pre}}=\frac{48\pi na_{s}\hbar^{2}}{m}=k_{\mathrm{B}}T^{\ast}$, giving $T^{\ast}\simeq54\mathrm{nK}$. We predict a quantum-thermal dynamical crossover at temperature $T^{\ast}\simeq54\mathrm{nK}$ that is below the BEC critical temperature $T_{C}\simeq92\mathrm{nK}$. This crossover marks a transition between two distinct dynamical classes for the post-quench relaxation, driven by a competition between quantum-fluctuation-dominated dynamics at low temperature and thermally excited defect-mediated relaxation at higher temperature. Below $T^{\ast}$, the growth of the spatial coherence is governed by bare quantum dispersion of the condensate. Above $T^{\ast}$, thermally excited noncondensed Bogoliubov quasiparticles scatter off phase fluctuations of the condensate, and two-body interactions control the growth of spatial coherence.

Underlying the relaxation toward equilibrium in isolated many-body ultracold-atom systems is quantum thermalization described by the eigenstate thermalization hypothesis (ETH)~\cite{RigolNat2008, DAlessioAP2016}. This thermalization relies on ergodicity on the constant-energy shell, where long-time observables follow microcanonical statistics. Accordingly, our previously discussed coherent spreading, which is linked to many-body entanglement scrambling during thermalization, naturally exhibits energy dependence. However, this energy dependence is weak at relatively low energies and becomes pronounced at relatively high energies. Since ETH and thermalization require non-integrable chaotic dynamics, this spreading should be absent in integrable settings~\cite{PolkovnikovRMP2011,DAlessioAP2016}. For pancake-shaped BECs, the governing two-dimensional nonlinear Schr\"odinger equation is nonintegrable and supports quantum chaos~\cite{ZhangPRL2004}, yet coherent spreading still follows the linear-growth law. By contrast, the one-dimensional nonlinear Schr\"odinger equation governing cigar-shaped BECs is exactly integrable~\cite{ZakharovJETP1972, Matveev1992, WuPRL2002, CazalillaRMP2011, Guo2025} and violates the ETH, such that linear-growth coherent spreading no longer holds in these systems. These theoretical predictions are corroborated by our extended numerical calculations and motivate further experimental investigations.

In summary, we investigate the relaxation dynamics of quenched ultracold Bose atoms and unambiguously identify a linear temporal growth law for coherence spreading, thereby establishing a coherence-spreading law. We demonstrate that the spreading speed exhibits universality to some extent, being insensitive to both initial momentum distributions and interparticle interaction strengths at relatively low initial energies. We further uncover that the spreading speed intrinsically depends on the system energies and is strictly governed by the system's integrability. Our ensemble-averaged GP trajectory approach provides a feasible way to study the relaxation and thermalization dynamics of far-from-equilibrium quantum many-body bosonic systems, and our theoretical predictions of a quantum-thermal dynamical crossover below the BEC critical temperature are readily accessible for future experimental verification.

\emph{Acknowledgments}---This work was supported by the Science Challenge Project (Grant No.~TZ2025017) and the National Natural Science Foundation of China (Grant No.~U2330401).

\bibliography{refs.bib}

\section*{End Matter}
The approximate constancy of $Z$ arises from the suppression of the net particle flux across $k_{\ast}$. For an isotropic spectrum within the continuum approximation, the nonzero-mode radial occupation density is $\rho_{N}(k,\Delta\tau)=\mathcal{V}k^{2}n_{k}(\Delta\tau)/(2\pi^{2})$, where $n_{k}(\Delta\tau)=\langle n_{\mathbf{k}}(\tau_{i}+\Delta\tau)\rangle$ and $k=|\mathbf{k}|$. We define $J_{N}(k,\Delta\tau)$ as the outward particle flux through $k$. Integrating the particle-conservation equation, $\partial_{\Delta\tau}\rho_{N}+\partial_{k}J_{N}=0$, over the low-momentum sector $k<k_{\ast}$ then yields $\partial_{\Delta\tau}Z=-J_{N}(k_{\ast},\Delta\tau)$.

To obtain the particle flux, we expand the field in the periodic box as $\phi=\sum_{\mathbf{k}}c_{\mathbf{k}}e^{i\mathbf{kx}}$ and obtain $i\partial_{\tau}c_{\mathbf{k}}=\omega_{k}c_{\mathbf{k}}+\frac{1}{2}\sum_{\mathbf{pqr}}c_{\mathbf{p}}^{\ast}c_{\mathbf{q}}c_{\mathbf{r}}\delta_{\mathbf{k}+\mathbf{p},\mathbf{q}+\mathbf{r}}$ from the dimensionless GP equation, where $\omega_{k}=k^{2}/2$ and the Kronecker delta enforces momentum conservation. The occupations $n_{\mathbf{k}}=|c_{\mathbf{k}}|^{2}$ satisfy $\sum_{\mathbf{k}}n_{\mathbf{k}}=1$ and obey $\partial_{\tau}n_{\mathbf{k}}=\text{Im}\sum_{\mathbf{pqr}}c_{\mathbf{k}}^{\ast}c_{\mathbf{p}}^{\ast}c_{\mathbf{q}}c_{\mathbf{r}}\delta_{\mathbf{k}+\mathbf{p},\mathbf{q}+\mathbf{r}}$. We apply the weak-wave kinetic approximation, in which the modes controlling the flux are treated as nearly linear waves exchanging particles through weak four-wave scattering. Within this approximation, phase correlations decay much faster than the occupations evolve, allowing memory effects to be neglected. Quasi-Gaussian factorization of the six-point correlations in the four-point evolution equations, followed by the long-time resonant limit, then gives $\partial_{\Delta\tau}n_{k}=I_{k}[n]$~\cite{DyachenkoPhyD1992, Nazarenko2011}. The continuum collision integral is $I_{k}[n] =\frac{\pi\mathcal{V}^{2}}{(2\pi)^{6}}\int\delta(\mathbf{k}+\mathbf{p}-\mathbf{q}-\mathbf{r})\delta(\omega_{k}+\omega_{p}-\omega_{q}-\omega_{r})\mathcal{F}_{kpqr}d^{3}\mathbf{p}d^{3}\mathbf{q}d^{3}\mathbf{r}$, where $\mathcal{F}_{kpqr}=n_{k}n_{p}n_{q}n_{r}(1/n_{k}+1/n_{p}-1/n_{q}-1/n_{r})$~\cite{ZhuPRE2022}. The radial density therefore obeys $\partial_{\Delta\tau}\rho_{N}(k,\Delta\tau)=\mathcal{V}k^{2}I_{k}[n(\Delta\tau)]/(2\pi^{2})$. Substituting this expression into the particle-conservation equation and integrating from $k$ to $k_{\mathrm{uv}}$, using the boundary condition $J_{N}(k_{\mathrm{uv}},\Delta\tau)=0$, we obtain $J_{N}(k,\Delta\tau)=\frac{\mathcal{V}}{2\pi^{2}}\int_{k}^{k_{\mathrm{uv}}}dpp^{2}I_{p}[n(\Delta\tau)]$, where $k_{\mathrm{uv}}$ denotes the maximum wave number retained in the classical-field model.

After energy pre-equilibration, we assume that the background modes and the resonant partners controlling the boundary flux remain close to a quasistationary Bose--Einstein distribution. In the high-occupation regime, this reduces to the RJ form $n_{k}^{\mathrm{RJ}}=\Theta/(\omega_{k}-\mu)$, where $\Theta$ and $\mu$ are the dimensionless effective temperature and chemical potential. Upon substituting the RJ distribution into $I_{p}$, the delta function in $I_{p}$ enforces $\omega_{k}+\omega_{p}=\omega_{q}+\omega_{r}$, giving $\mathcal{F}_{kpqr}\simeq0$ and hence $I_{p}[n^{\mathrm{RJ}}]\simeq0$ over the band entering the flux integral~\cite{NewellARFM2011,Nazarenko2011}. The flux integral then gives $J_{N}(k_{\ast},\Delta\tau)\simeq0$, and particle conservation yields $Z(\Delta\tau)\simeq Z(0)$.

\end{document}